\documentclass{aa}

\usepackage{graphicx}
\usepackage{txfonts}
\usepackage{multirow}
\begin{document}

\title{Is cosmic acceleration proven by local cosmological probes?}

\titlerunning{Is cosmic acceleration proven by local cosmological probes?}

\author{I. Tutusaus,
  \inst{1,2}\fnmsep\thanks{\email{isaac.tutusaus@irap.omp.eu}}
  B. Lamine, \inst{1,2} A. Dupays, \inst{1,2} \and A. Blanchard
  \inst{1,2} }

\institute{Universit\'e de Toulouse, UPS-OMP, IRAP, F-31400 Toulouse,
  France \and CNRS, IRAP, 14, avenue Edouard Belin, F-31400 Toulouse,
  France\\
}

\authorrunning{I. Tutusaus et al.}
\date{Received -; accepted -}

% \abstract{}{}{}{}{} 
% 5 {} token are mandatory
 
\abstract
{The cosmological concordance model ($\Lambda$CDM) matches the
  cosmological observations exceedingly well. This model has become the
  standard cosmological model with the evidence for an accelerated
  expansion provided by the type Ia supernovae (SNIa) Hubble diagram.
  However, the robustness of this evidence has been addressed
  recently with somewhat diverging conclusions.
}
{The purpose of this paper is to assess the robustness of the
  conclusion that the Universe is indeed accelerating if
  we rely only on low-redshift ($z\lesssim2$) observations, that is to
  say with SNIa, baryonic acoustic oscillations, measurements of
  the Hubble parameter at different redshifts, and measurements of the
  growth of matter perturbations.
}
{We used the standard statistical procedure of minimizing
  the $\chi^2$ function for the different probes to quantify
  the goodness of fit of a model for both $\Lambda$CDM and a simple
  nonaccelerated low-redshift
  power law model. In this analysis, we do not assume that supernovae
  intrinsic luminosity is independent of the redshift, which has been a fundamental assumption
  in most previous studies that cannot be tested.
}
{We have found that, when SNIa intrinsic luminosity is not assumed to be
  redshift independent, a nonaccelerated low-redshift power law model
  is able to fit the low-redshift background data as
  well as, or even slightly better, than $\Lambda$CDM. When measurements of the growth of structures are
  added, a nonaccelerated low-redshift power law model still
  provides an excellent fit to the data for all the luminosity evolution models considered.  }
{Without the standard assumption that supernovae intrinsic luminosity is independent of
  the redshift, low-redshift probes are consistent with a nonaccelerated universe.
%not sufficient to rule out power-law  decelerated models of the expansion of the Universe.
}


\maketitle
%
%-------------------------------------------------------------------

\section{Introduction}
The cosmological concordance model ($\Lambda$CDM) framework offers a
simple description of the properties of the Universe reproducing
noticeably well a wealth of high quality observations. However, we do
not know the true nature of the dark components of the $\Lambda$CDM
model, which form about 95\% of the energy content of the
Universe. The $\Lambda$CDM model has become the standard cosmological model 
with the evidence for an accelerated expansion provided by the type Ia supernovae (SNIa) Hubble diagram
%is mainly founded on the accelerated
%expansion of the Universe at present, which was originally inferred
%from type Ia supernovae (SNIa) data
[\cite{Riess,Perlmutter}]. However, there has recently been an
important discussion in the literature concerning the ability of SNIa
data alone to prove the accelerated expansion of the Universe
[\cite{Sarkar,Shariff,Rubin,Ringermacher}].

In this paper %work we focus on the problem 
we examine whether the accelerated nature of the expansion can be firmly established based not only on SNIa
data, but also on the other low-redshift cosmological probes: the baryon
acoustic oscillations (BAO), the Hubble parameter as a function of
the redshift ($H(z)$), and measurements of the growth of structures ($f\sigma_8(z)$). Moreover, 
%given the recent discussion about thestatistical tools needed to analyze SNIa data [\cite{Sarkar,Rubin}],
we do not assume that the intrinsic luminosity of supernovae is
independent of the redshift; therefore, we consider a nuisance parameter accounting for some luminosity
evolution of SNIa with redshift, and we consider a large variety
of luminosity evolution models to be as general as
possible. We discard high-redshift data, such as cosmic microwave
background (CMB), because they are sensitive to the early Universe physics and
because our goal is just to assess whether measurements of the local Universe are sufficient to
prove the accelerated expansion of the Universe, which, at least in the
standard cosmological model, appears at low redshift. In order to do this, we
consider a simple nonaccelerated model based on
a power law cosmology [\cite{ref3,ref6,ref4,ref5,ref7,ref19,Tutusaus}],
but we focus here on a cosmological model that behaves like a power law
cosmology only at low redshift, while the model behavior at
high redshift is irrelevant for the cosmological probes considered in
this work. We denote this model by NALPL (nonaccelerated local power
law). The
power law cosmology states that the scale factor, $a(t)$, evolves
proportionally to some power of  time, $a(t)\propto t^n$. Since we are
interested in proving the acceleration of the Universe, we limit $n
\leq 1$ to deal with a nonaccelerating universe at late time.

In Sec.\,\ref{sec2} we briefly describe the cosmological models under
consideration. In Sec.\,\ref{sec3} we present the statistical tool
used to determine the goodness of fit of the models to the data. %between themselves. 
In Sec.\,\ref{sec4} we
present the low-redshift probes used in this study: SNIa, BAO, $H(z)$
and $f\sigma_8(z)$, as well as the different luminosity evolution models
considered. We provide the results in Sec.\,\ref{sec5} and we conclude
in Sec.\,\ref{sec6}.

%--------------------------------------------------------------------
\section{Models}\label{sec2}
In this section we present the two models used in this analysis: the
$\Lambda$CDM model and the nonaccelerated low-redshift power law
cosmology (NALPL). 

\subsection{Cosmological concordance model}
The flat $\Lambda$CDM model is the current standard model in cosmology
thanks to its ability to fit the main cosmological data,
SNIa [\cite{SNIa}], BAO [\cite{BAO}], and the CMB [\cite{CMB}]. This model assumes
a flat Robertson-Walker metric together with Friedmann-Lema\^{i}tre
dynamics leading to the comoving angular diameter distance,

\begin{equation}
r(z)=c\int_0^z\frac{\text{d}z'}{H(z')}\,,
\end{equation}
and the Friedmann-Lema\^{i}tre equation,

\begin{equation}
H(z)=H_0\sqrt{\Omega_r(1+z)^4+\Omega_m(1+z)^3+(1-\Omega_r-\Omega_m)}\,,
\end{equation}
where $H_0$ is the Hubble constant and $\Omega_i$ is the energy
density parameter of the fluid $i$. We follow \cite{CMB} in computing
the radiation contribution as
\begin{equation}
\Omega_r=\Omega_{\gamma}\left[1+N_{\text{eff}}\frac{7}{8}\left(\frac{4}{11}\right)^{4/3}\right]\,,
\end{equation}
where $\Omega_{\gamma}$ represents the photon contribution and is
given by
\begin{equation}
\Omega_{\gamma}=4\cdot 5.6704\times
10^{-8}\frac{T_{\text{CMB}}^4}{c^3}\frac{8\pi G}{3H_0^2}\,.
\end{equation}

We fix\footnote{We checked that small variations on these
  parameters do not modify the results.} the effective number of
neutrino-like relativistic degrees of freedom, $N_{\text{eff}}=3.04$
[\cite{CMB}], $H_0=67.74\,\text{km}\,\text{s}^{-1}\,\text{Mpc}^{-1}$
[\cite{CMB}], and the temperature of the CMB today,
$T_{\text{CMB}}=2.725$ [\cite{TCMB}]. For simplicity, we fix $H_0$ only for the
radiation contribution in the $\Lambda$CDM model. This parameter
is left free in the rest of the work.

\subsection{Nonaccelerated low-redshift power law model}
The local power law cosmology is formulated in such a way that the scale factor is related to
the proper time through a power law relation,
\begin{equation}
a(t)=\left(\frac{t}{t_0}\right)^n\,,
\end{equation}
at late time. The Friedmann-Lema\^{i}tre equation reads
\begin{equation}
H(z)=H_0(1+z)^{1/n}\,,
\end{equation}
so that the comoving angular diameter distance yields
\begin{equation}
r(z)=\frac{c}{H_0}\times\left\{
\begin{array}{cc}
\frac{(1+z)^{1-1/n}-1}{1-1/n}, & n\neq 1\,,\\
\ln(1+z), & n=1\,.
\end{array}\right.
\end{equation}

We limit the cosmological parameter $n$ to be smaller or equal to 1 to have a nonaccelerated universe.

\section{Method}\label{sec3}

\begin{figure*}
\begin{center}
\includegraphics[scale=.45]{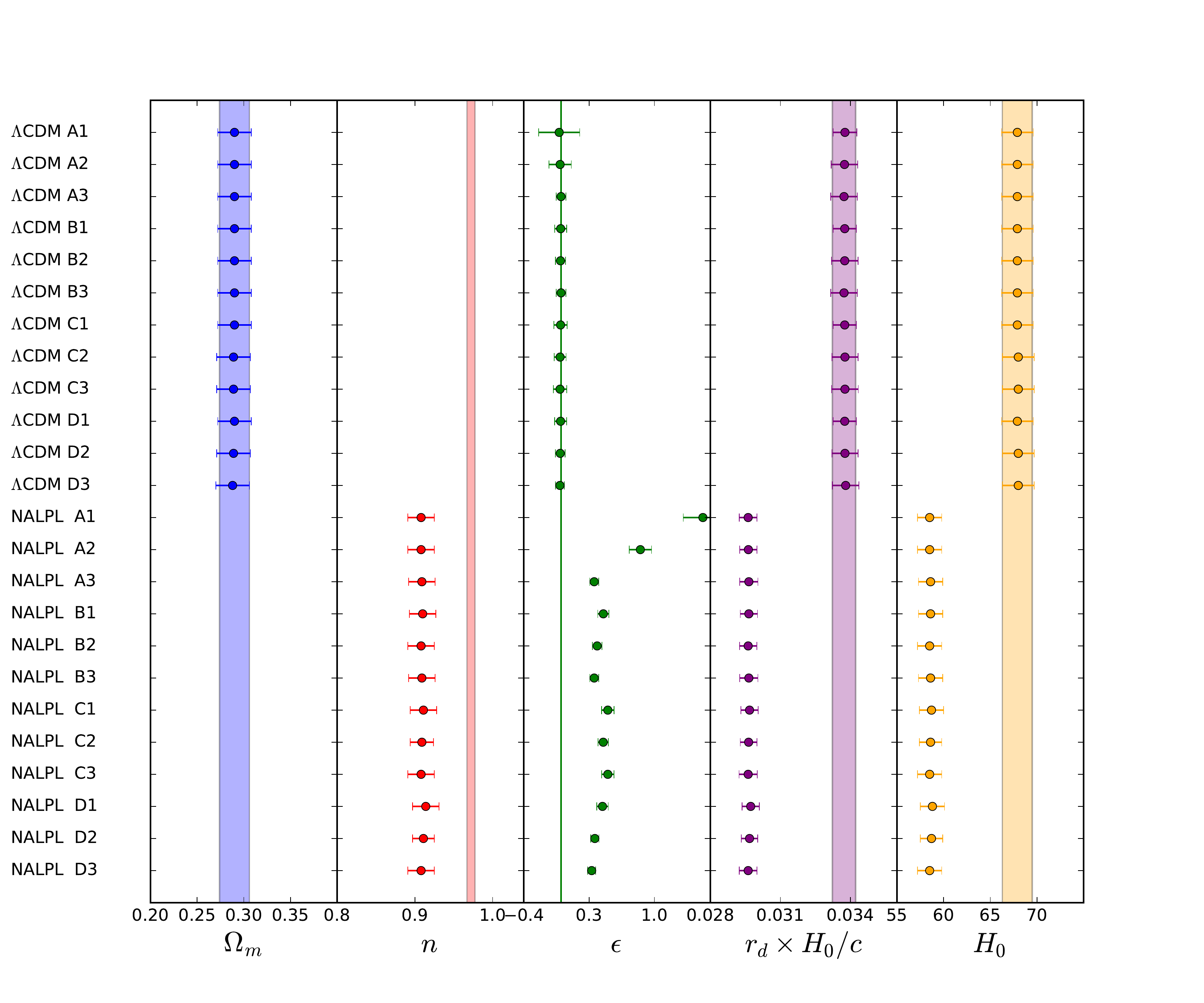}\\
\includegraphics[scale=.45]{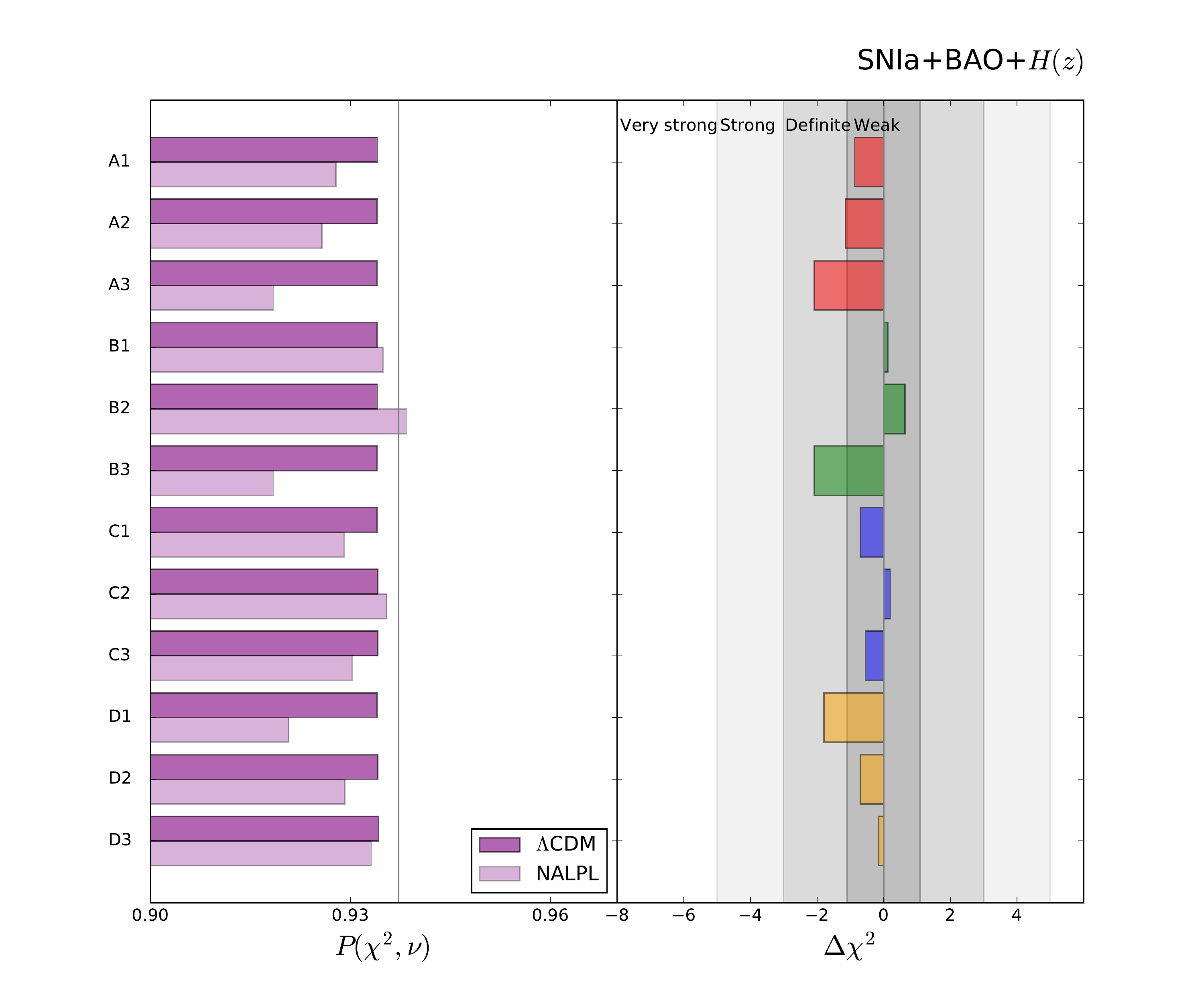}
\caption{Results obtained from low-redshift background probes. Top panel: Best-fit values
  for $\Omega_m$, $n$, $\epsilon$, $r_d\times H_0/c,$ and $H_0$ parameters for all the
  cosmological and luminosity evolution models under consideration. The
  values for these parameters when no luminosity evolution is allowed
  are represented with bands as a reference. Bottom panel: Goodness of
  fit statistics and difference of the $\chi^2$ values, $\Delta
  \chi^2=\chi^2_{\Lambda \text{CDM}}-\chi^2_{\text{NALPL}}$, for the luminosity
  evolution models under study. The vertical solid line in the left plot
  illustrates the goodness of fit statistics for the standard $\Lambda$CDM
  imposing no luminosity  evolution. The various gray bands
  in the right plot show the strength
  of $\Delta \chi^2$ given by the Jeffrey scale (see the text for
  details). }\label{fig1}
\end{center}
\end{figure*}

In this section we review the statistical tool used to determine
the ability of a model to fit the cosmological data.

To quantify the goodness of a fit we minimize the $\chi^2$ function
given by
\begin{equation}
\chi^2=(\textbf{u}-\textbf{u}_{\text{data}})^TC^{-1}(\textbf{u}-\textbf{u}_{\text{data}})\,,
\end{equation}
where \textbf{u} stands for the model prediction, while
$\textbf{u}_{\text{data}}$ and $C$ hold for the observables and their
covariance matrix, respectively. In order to perform the minimization
and find the errors on the parameters, we use the MIGRAD application
from the \texttt{iminuit} Python
package\footnote{\url{https://github.com/iminuit/iminuit}}. This package is the
Python implementation of the former MINUIT Fortran code
[\cite{minuit}].

We also compute the probability that a higher value for the $\chi^2$
occurs for a fit with $\nu=N-k$ degrees of freedom, where $N$ is the
number of data points and $k$ is the number of parameters of the
model,
\begin{equation}\label{prob}
P(\chi^2,\nu)=\frac{\Gamma\left(\frac{\nu}{2},\frac{\chi^2}{2}\right)}
{\Gamma\left(\frac{\nu}{2}\right)}\,,
\end{equation}
where $\Gamma(t,x)$ is the upper incomplete gamma function and
$\Gamma(t)=\Gamma(t,0)$ the complete gamma function. We use this value
as a goodness of fit statistic. A probability
close to 1 indicates that it is likely to obtain higher $\chi^2$
values than the minimum found, pointing to a good fit by the
model.

When combining probes, we minimize the sum of the individual $\chi^2$
functions, i.e., we assume that we are dealing with statistically
independent probes. Equation\,(\ref{prob})
is only valid for $N$ data points coming from $N$ independent random
variables with Gaussian distributions. In this work we use the
correlations within probes; therefore, our data points no longer
come from independent Gaussian random variables. However, it has been
shown in \cite{Tutusaus}, through Monte Carlo simulations, that the
impact of the correlations we are dealing with is negligible in
Eq.\,(\ref{prob}).

\section{Data samples}\label{sec4}

%\begin{figure*}
%\begin{center}
%\includegraphics[scale=.45]{fig3a.pdf}\\
%\includegraphics[scale=.45]{fig3b.pdf}
%\caption{Results obtained from SNIa and BAO data. Top panel: Best-fit
  %values for $\Omega_m,\,n,\,\epsilon$ and $r_d\times H_0/c$
  %parameters for all the models under consideration. Bottom panel:
  %Model comparison criteria values for all the models with
  %$\Lambda$CDM with and without evolution as reference models (see Fig.\,\ref{fig1}).}\label{fig3}
%\end{center}
%\end{figure*}

%\begin{figure}
%\begin{center}
%\includegraphics[scale=.45]{fig4.pdf}
%\caption{Luminosity evolution function as a function of the redshift
  %for the three most preferred models against the standard
  %$\Lambda$CDM using SNIa and BAO data: power law B1, C1 and D2.}\label{fig4}
%\end{center}
%\end{figure}

%\begin{figure*}
%\begin{center}
%\includegraphics[scale=.45]{fig5a.pdf}\\
%\includegraphics[scale=.45]{fig5b.pdf}
%\caption{Results obtained from SNIa, BAO and $H(z)$ data. Top panel: Best-fit
  %values for $\Omega_m,\,n,\,\epsilon,\,r_d\times H_0/c$ and $H_0$ parameters for all the models under consideration. Bottom panel:
 % Model comparison criteria values for all the models with
  %$\Lambda$CDM with and without evolution as reference models (see Fig.\,\ref{fig1}).}\label{fig5}
%\end{center}
%\end{figure*}

In this section we present the low-redshift probes, SNIa, BAO, $H(z),$
and $f\sigma_8(z)$, and the specific data samples used in this work.

\subsection{Type Ia supernovae}
Type Ia supernovae are considered standardizable candles useful to
measure cosmological distances and break some degeneracies present
in other probes, providing us precise cosmological measurements. The
standard observable used in SNIa measurements is the distance modulus,
\begin{equation}
\mu(z)=5\log_{10}\left(\frac{H_0}{c}d_L(z)\right)\,,
\end{equation}
where $d_L(z)=(1+z)r(z)$ is the luminosity distance.

The standardization of SNIa is based on empirical observation that
they form a homogeneous class whose variability can be characterized
by two parameters [\cite{SNIaparams}]: the time stretching of the
light curve $(X_1)$ and the supernova color at maximum brightness
$(C)$. In this work we use the joint light-curve analysis for SNIa
from \cite{SNIa}. Given the assumption of the authors that supernovae
with identical color, shape, and galactic environment have on average
the same intrinsic luminosity for all redshifts, the distance modulus
can be expressed as
\begin{equation}
\mu_{\text{obs}}=m_B^*-(M_B-\alpha X_1+\beta C)\,,
\end{equation}
where $m_B^*$ corresponds to the observed peak magnitude in the B-band
rest-frame and $\alpha$ and $\beta$ are nuisance parameters related to
the time stretching and the supernova color, respectively. The $M_B$
nuisance parameter takes into account the supernova dependence on host
galaxy properties and is given by
\begin{equation}
M_B=\left\{
\begin{array}{cl}
M_B^1\,, & \text{if}\,\, M_{\text{stellar}} < 10^{10} M_{\odot}\,,\\
M_B^1+\Delta M\,, & \text{otherwise}\,,
\end{array}\right.
\end{equation}
where $M_b^1$ and $\Delta M$ are two extra nuisance parameters.

Concerning the errors and correlations of the measurements, we use the
covariance matrix provided in \cite{SNIa}, where several statistical
and systematic uncertainties have been considered, such as the error
propagation of the light-curve fit uncertainties, calibration, light-curve model, bias correction, mass step, dust
extinction, peculiar velocities, and contamination of non-type
Ia supernovae.  It is important to stress that this covariance matrix
depends on the $\alpha$ and $\beta$ nuisance parameters; therefore,
when performing a minimization we recompute the covariance matrix at
each step.

Since we do not assume that the intrinsic luminosity of SNIa is
independent of the redshift, we consider an extra nuisance term,
$\Delta m_{\text{evo}}(z)$, accounting for a possible evolution of the
supernovae luminosity with the redshift,
\begin{equation}
\mu_{\text{obs}}=m_B^*-(M_B-\alpha X_1+\beta C+\Delta m_{\text{evo}}(z))\,.
\end{equation}

Different phenomenological models for $\Delta m_{\text{evo}}(z)$ can
be found in the literature [see for example
\cite{Drell,Linder2006,Nordin,Ferr2009,SNIaevII}]. In the absence of
any clear physics governing this evolution, one stays at a
phenomenological level and considers a bunch of different models. We
can embed all the models studied in this paper into four categories,
which are summarized in table~\ref{table:evol model}. These categories all possess two
parameters, $\epsilon$ and $\delta$.

Model B is equivalent to model 2 in \cite{SNIaevII}, while model C is
a generalization of model 1 in \cite{SNIaevII}. Model D is a
generalization of \cite{Ferr2009}. Models C and D were initially
motivated from a parameterization of the intrinsic luminosity
$\mathcal{L}\rightarrow \mathcal{L}(1+z)^{-\epsilon}$ [\cite{Drell}],
while models A and B are more general to study the contribution of
powers in $z$ to the results. 

\begin{table}
  \caption{Different evolution models for SNIa considered in this
  paper. All models have two different parameters, $\epsilon$ and
  $\delta$.}
  \centering
\begin{tabular}[c]{lll}
Model&$\Delta m_{\text{evo}}(z)$&Reference\\\hline \hline
A&$\epsilon\left[(1+z)^{\delta}-1\right]$&-\\[2mm]
B&$\epsilon z^{\delta}$&\cite{SNIaevII}\\[2mm]
C&$\epsilon\left[\ln(1+z)\right]^{\delta}$&\cite{SNIaevII}\\
D&$\displaystyle\epsilon\left(\frac{t_0-t(z)}{t_0-t(1)}\right)^{\delta}$&\cite{Ferr2009}\\
\end{tabular}
\label{table:evol model}
\end{table}

% : model A,

% \begin{equation}\label{modela}
% \Delta m_{\text{evo}}(z)=\epsilon\left[(1+z)^{\delta}-1\right]\,,
% \end{equation}
% model B [equivalent to model 2 in \cite{SNIaevII}],

% \begin{equation}\label{modelb}
% \Delta m_{\text{evo}}(z)=\epsilon z^{\delta}\,,
% \end{equation}
% model C [generalization of model 1 in \cite{SNIaevII}],

% \begin{equation}\label{modelc}
% \Delta m_{\text{evo}}(z)=\epsilon\left[\ln(1+z)\right]^{\delta}\,,
% \end{equation}
% and model D [generalization of \cite{Ferr2009}],

% \begin{equation}\label{modeld}
% \Delta m_{\text{evo}}(z)=\epsilon\left(\frac{t_0-t(z)}{t_0-t(1)}\right)^{\delta}\,.
% \end{equation}

Let us observe that when $\delta\rightarrow 0$,
$\Delta m_{\text{evo}}(z)$ becomes strongly degenerate with
$M_B^1$. Therefore, to avoid these kinds of parameter
degeneracies, we consider three
submodels fixing $\delta=0.3,\,0.5,$ and $1$. We denote these
submodels  A1, A2, A3, B1, B2, B3, and so on. A lower
$\delta$ power contribution models a luminosity evolution dominant at low
redshift, while a higher $\delta$ power contribution leads to a luminosity
evolution dominating at high redshift. 

When using SNIa data, the set of nuisance parameters considered is
$\{\alpha,\,\beta,\,M_B^1,\,\Delta M,\epsilon\}$. We consider
$\Omega_m$ and $n$ as cosmological parameters for the $\Lambda$CDM and
the NALPL model, respectively.

\subsection{Baryonic acoustic oscillations}
\begin{figure*}
\begin{center}
\includegraphics[scale=.45]{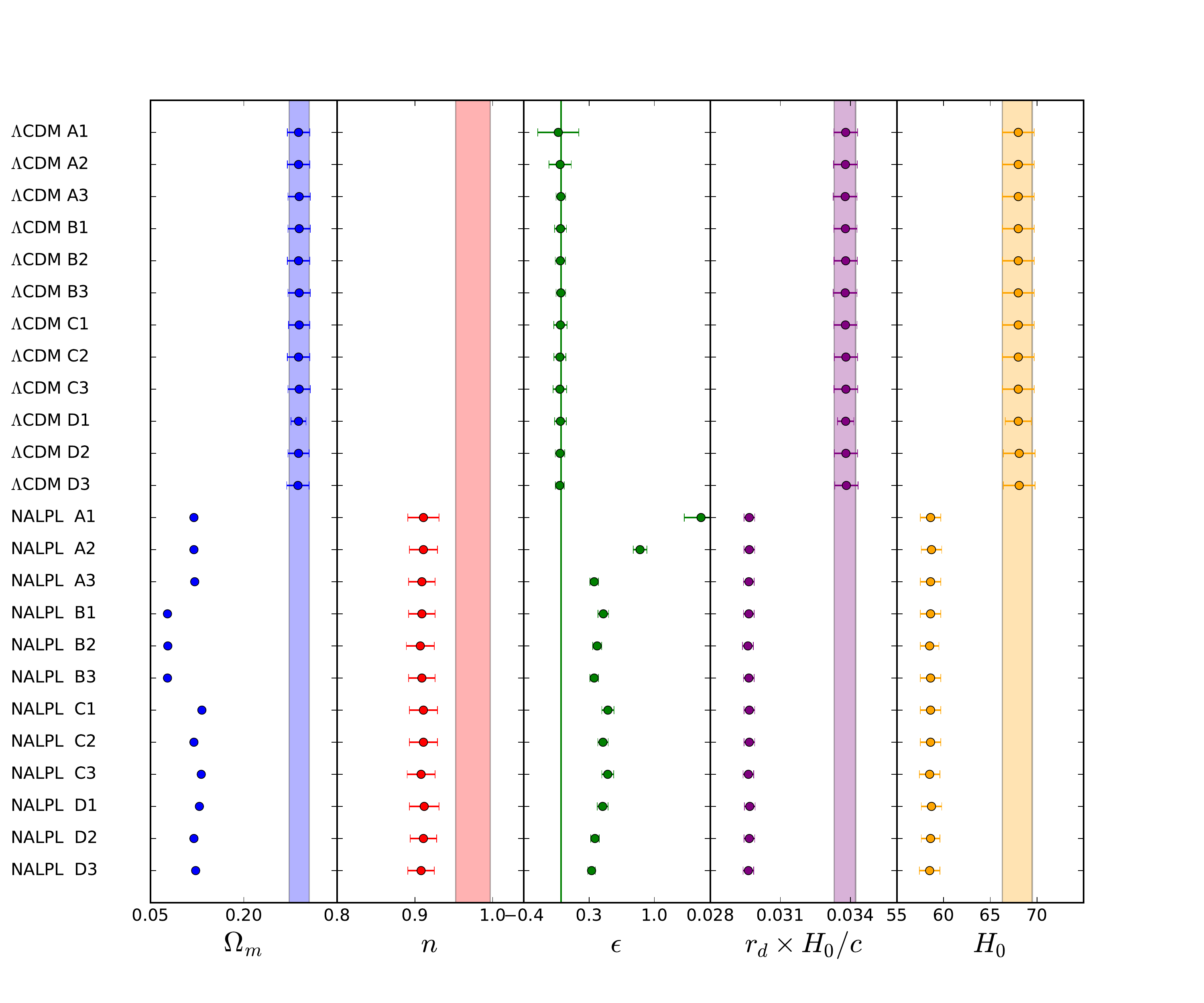}\\
\includegraphics[scale=.45]{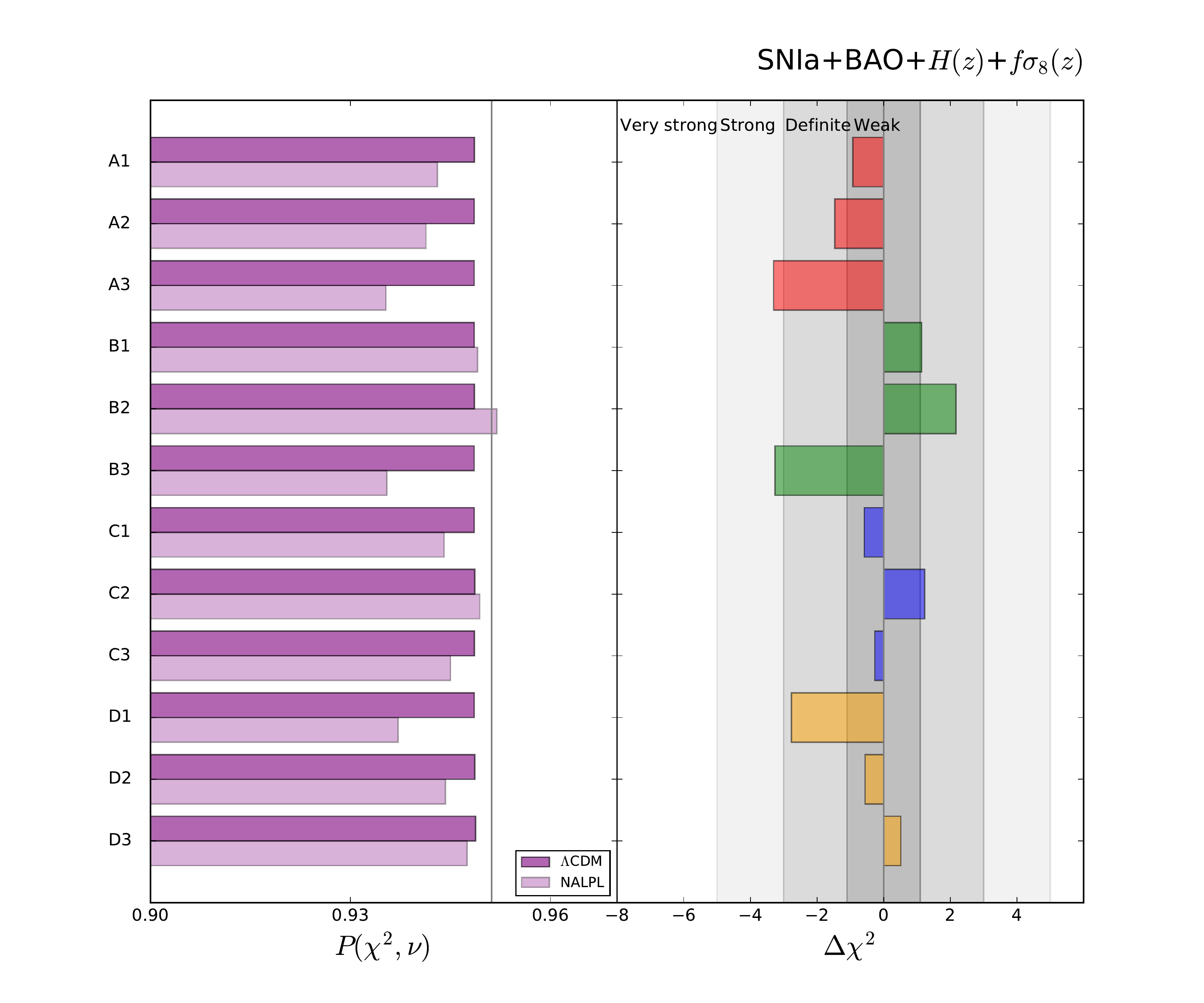}
\caption{Results obtained from low-redshift background probes combined
  with measurements of the growth of matter perturbations. Top panel: Best-fit values
  for $\Omega_m$, $n$, $\epsilon$, $r_d\times H_0/c$, and $H_0$ parameters for all the
  cosmological and luminosity evolution models under consideration. The
  values for these parameters when no luminosity evolution is allowed
  are represented with bands as a reference. Bottom panel: Goodness of
  fit and $\Delta\chi^2=\chi^2_{\Lambda\text{CDM}}-\chi^2_{\text{NALPL}}$ values for the luminosity
  evolution models under study (see Fig.\,\ref{fig1}). }\label{fig2}
\end{center}
\end{figure*}

The baryonic acoustic oscillations are the regular and periodic
fluctuations of visible matter density in large-scale structure. They
are characterized by the length of a standard ruler, generally denoted
by $r_d$. In the $\Lambda$CDM model, the BAO come from sound waves
propagating in the early Universe and $r_d$ is equal to the comoving
sound horizon at the redshift of the baryon drag epoch, 
\begin{equation}\label{rd}
r_d=r_s(z_d)=\int_{z_d}^{\infty}\frac{c_s(z)\,\text{d}z}{H(z)}\,,
\end{equation}
where $z_d\approx 1060$ and $c_s(z)$ is the sound velocity as a
function of the redshift.  \cite{Verde} have shown that
models differing from $\Lambda$CDM may have a
value for $r_d$ that is not compatible with $r_s(z_d)$. Moreover, the integral in
Eq.\,(\ref{rd}) is divergent for $n\geq 2/3$ power
law cosmologies. According to this, and in order not to delve into
early Universe physics, we consider $r_d$ as a free parameter.

In this work we use isotropic and anisotropic measurements of the
BAO. The distance scale used for isotropic measurements is given by
\begin{equation}
D_V(z)\equiv\left(r^2(z)\frac{cz}{H(z)}\right)^{1/3}\,,
\end{equation}
while for the radial and transverse measurements of the anisotropic
BAO the distance scales are $r(z)$ and $c/H(z)$, respectively.

We use the values provided by 6dFGS [\cite{BAO1}], SDSS - MGS
[\cite{BAO2}], BOSS - CMASS, and LOWZ samples DR11 [\cite{BAO,BAO3}]
and BOSS - Ly$\alpha$ forest DR11 [\cite{BAO4,BAO5}]. We consider a
correlation coefficient of $0.52$ for the CMASS measurements and of
$-0.48$ for the Ly-$\alpha$ measurements, while we assume the rest of
the measurements to be uncorrelated. In order to take into account the
non-Gaussianity of the BAO observable likelihoods far from the peak,
we follow \cite{BAONG} by replacing the usual
$\Delta \chi^2_G=-2\ln\mathcal{L}_G$ for a Gaussian-likelihood
observable by
\begin{equation}
\Delta \chi^2=\frac{\Delta \chi^2_G}{\sqrt{1+\Delta\chi^4_G
\left(\frac{S}{N}\right)^{-4}}}\,,
\end{equation}
where $S/N$ is the detection significance, in units of $\sigma$, of
the BAO feature. We follow \cite{Tutusaus} in considering a detection
significance of $2.4\sigma$ for 6dFGS, $2\sigma$ for SDSS-MGS,
$4\sigma$ for BOSS-LOWZ, $6\sigma$ for BOSS-CMASS, and $4\sigma$ for
BOSS-Ly-$\alpha$ forest.

When using BAO data, we add the following parameters to our set:
$\{r_d\times H_0/c,\,\Omega_m,\,n\}$. The latter two only apply for
the $\Lambda$CDM model and NALPL model, respectively. No
nuisance parameters are added.

\subsection{Hubble parameter $H(z)$}
There are two main methods to measure the evolution of the Hubble
parameter with respect to the redshift: the so-called differential age method
[\cite{Hz1}] and a direct measure of $H(z)$ using radial BAO
information [\cite{Gaztanaga}]. A detailed discussion on the
systematic uncertainties of these methods can be found in
\cite{Hz2}. In this work we use the compilation of independent $H(z)$
measurements from \cite{refH1,refH2,refH3,refH4,refH5,refH6,refH7} provided in \cite{Hz3}. When using $H(z)$ measurements, we add
the $H_0$ parameter to our set of cosmological parameters under
consideration. No nuisance parameters are added.

\subsection{Growth rate}

The measurements of the growth rate of matter perturbations offer an additional
constraint on cosmological models. Their value depends on the theory
of gravity used and it is well known that identical background
evolution can lead to differente growth rates [\cite{EFT}]. Let us first define the linear growth factor of matter perturbations as the ratio between
the linear density perturbation and the energy density,

\begin{equation}
D\equiv
\delta \rho_m/\rho_m\,. 
\end{equation}

We can then derive the standard second order differential equation for
the linear growth factor [\cite{Peebles}]

\begin{equation}
\ddot{D}+2H\dot{D}-4\pi G \rho_m D=0\,,
\end{equation}
where the dot stands for differentitation over the cosmic
time. Neglecting second order corrections, this differential equation
can be rewritten with derivatives over the scale factor [\cite{Dodelson}]

\begin{equation}\label{dsolve}
D''(a)+\left[\frac{3}{a}+\frac{H'(a)}{H(a)}\right]D'(a)-\frac{3}{2}\Omega_m\frac{H_0^2}{H^2(a)}\frac{D(a)}{a^5}=0\,.
\end{equation}

This expression is only valid if we
assume that dark energy cannot be perturbated and does not interact
with dark matter. Once we obtained $D$ by solving numerically
Eq.\,(\ref{dsolve}), we can compute the growth rate as

\begin{equation}
f\equiv \frac{\text{d}\ln D}{\text{d}\ln a}\,,
\end{equation}
and the observable weighted growth rate, $f\sigma_8$, as

\begin{equation}
f\sigma_8(z)=f(z)(K \cdot   D(z))\,,
\end{equation}
where $K$ is a normalization factor accounting for the root mean
square mass fluctuation amplitude on scales of $8h^{-1}$Mpc at
redshift $z=0$, $\sigma_8$, and the normalization of the growth
factor.  We do not use the observed value for $\sigma_8$
since it requires some assumptions about the early Universe
physics. Instead, we let the low-redshift observations choose the
preferred value for this normalization.

In this work we use the measurements of $f\sigma_8(z)$ from
\cite{fsig8ref1,fsig8ref2,fsig8ref3,refH6,fsig8ref5} provided in
\cite{fsig8data}, together with their correlations. When using these
measurements we add the $K$ parameter and $\Omega_m$, for the NALPL model, to our set of parameters under consideration.

\section{Results}\label{sec5}

\begin{figure*}
\begin{center}
\includegraphics[scale=.55]{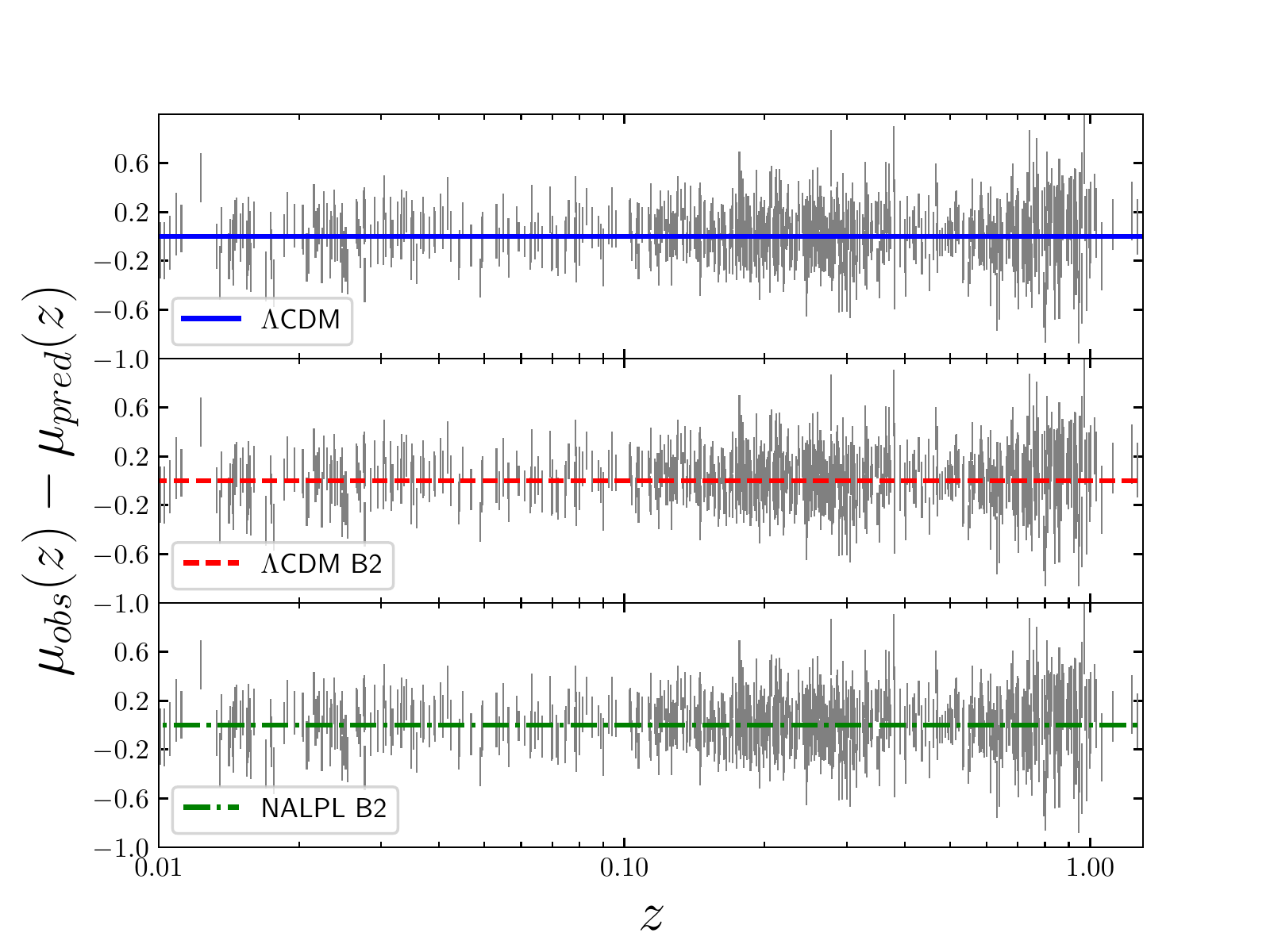}\,\includegraphics[scale=.55]{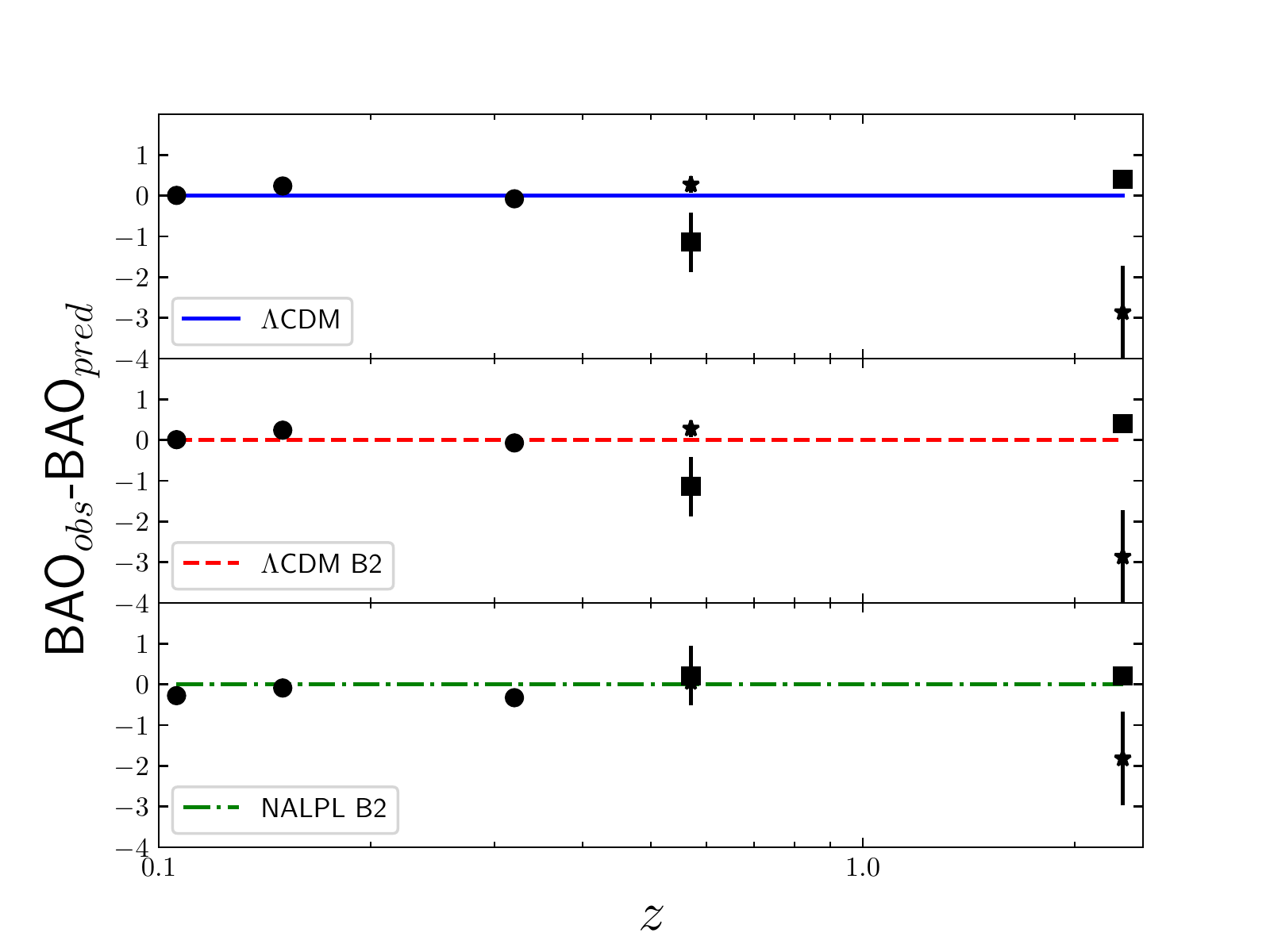}\\
\includegraphics[scale=.55]{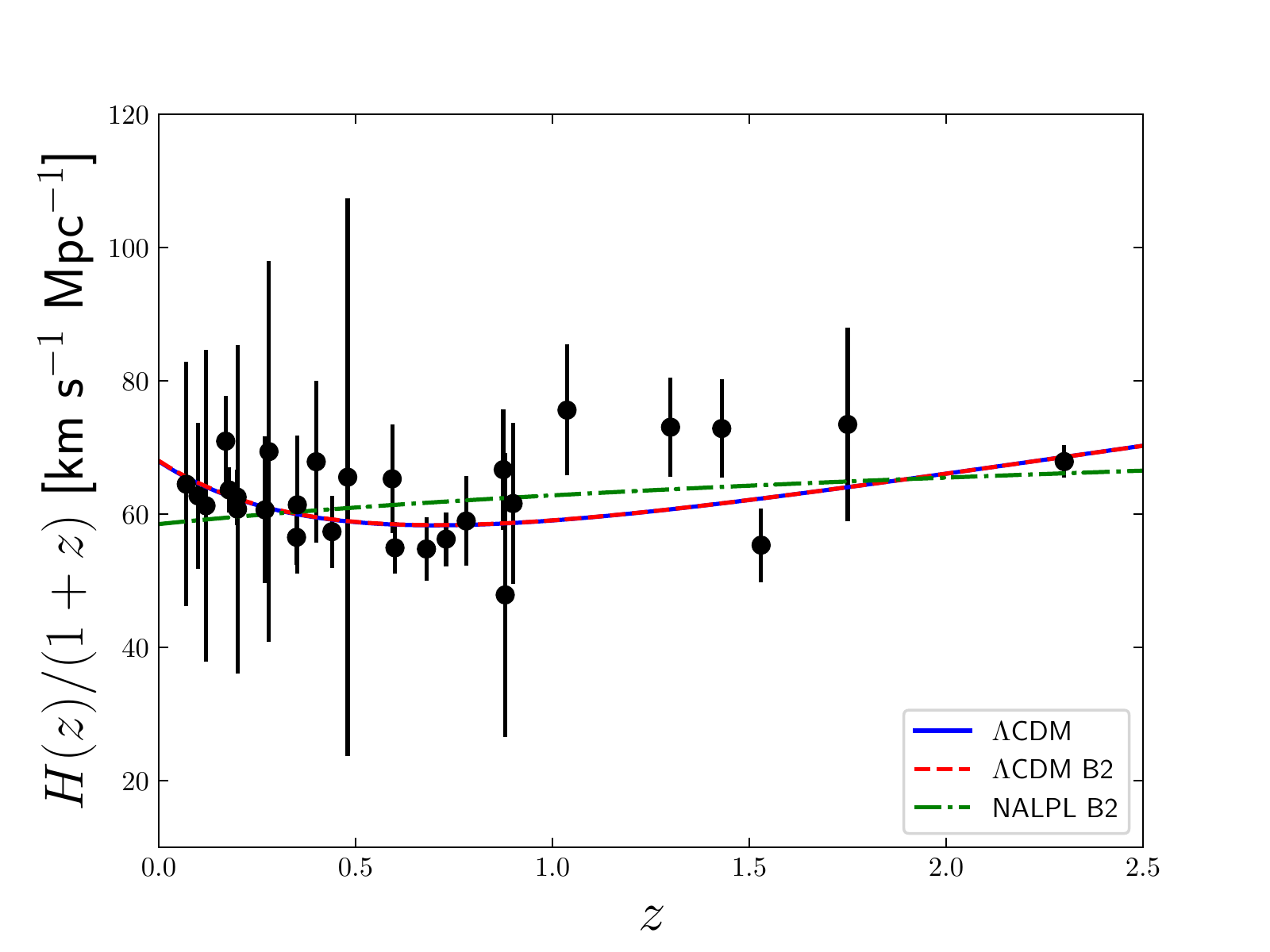}\,\includegraphics[scale=.55]{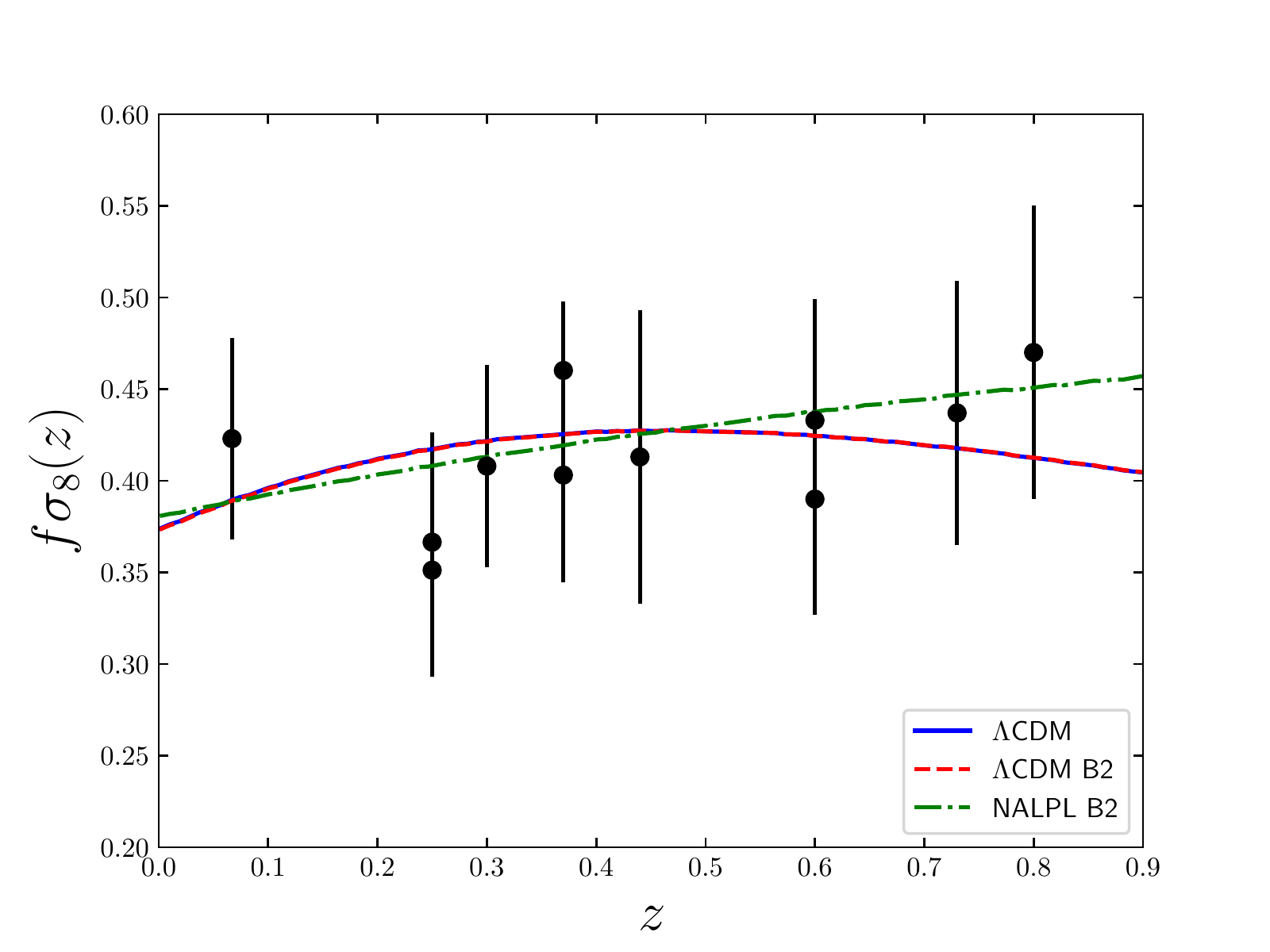}
\caption{Model predictions vs. observations for the best luminosity
evolution model, $\Lambda$CDM B2 and NALPL B2 (together with the
standard $\Lambda$CDM prediction, for illustrative purposes) for all
the cosmological probes considered. In each prediction, we used the best-fit
values obtained from the global fit. Top left:
Residuals of SNIa data with respect to the corresponding model. Top
right: Residuals of BAO data with respect to the corresponding
model. The isotropic measurements of the BAO are represented with a
circle and their observable is $D_V(z)/r_d$, while the stars stand for
the radial measurements with observable $r(z)/r_d$ and the squares
stand for the transverse measurements with observable
$c/(H(z)r_d)$. Bottom left: Measurements of $H(z)/(1+z)$ together with the model
predictions. Bottom right: Measurements of $f\sigma_8(z)$ and the
different model predictions.}\label{fig3}
\end{center}
\end{figure*}

%\begin{figure}
%\begin{center}
%\includegraphics[scale=.45]{fig2.pdf}
%\caption{Luminosity evolution function as a function of the redshift
  %for the three most preferred models against the standard
  %$\Lambda$CDM using SNIa+BAO+$H(z)$ data: B1, B2 and C2.}\label{fig2}
%\end{center}
%\end{figure}

%\begin{figure}
%\begin{center}
%\includegraphics[scale=.45]{../../Plots/BAO_plot_SNIa+BAO+Hz.pdf}
%\caption{}\label{fig6}
%\end{center}
%\end{figure}

%\begin{figure}
%\begin{center}
%\includegraphics[scale=.45]{../../Plots/h_of_z_SNIa+BAO+Hz.pdf}
%\caption{}\label{fig7}
%\end{center}
%\end{figure}

We present the results of this work in two different steps. In the first place we
focus on low-redshift background probes, namely SNIa, BAO, and $H(z)$,
and in the second step we add the measurements of the growth of matter perturbations.

The results obtained from low-redshift background probes only are
presented in Fig.\,\ref{fig1}. In the top panel we show the best-fit
values obtained for the $\Omega_m$, $n$, $r_d\times H_0/c$ and $H_0$ cosmological parameters,
as well as the $\epsilon$ nuisance parameter for the different cosmological
and luminosity evolution models under study. The blue region of the
left panel corresponds to the value, and the error, obtained for
$\Omega_m$ in the standard $\Lambda$CDM case, i.e., with no luminosity
evolution. We can observe that all the obtained values for $\Omega_m$
are completely compatible with the standard $\Lambda$CDM value.

In the second panel we plot the exponent $n$ of NALPL for each model, together with a colored band corresponding
to the allowed values when no evolution is imposed. We can observe
that all the obtained values are compatible with a slightly lower
value than in the no evolution case.

In the third panel we present the nuisance parameter associated with the
luminosity evolution for all the models under consideration. As
expected, all the $\Lambda$CDM models are perfectly compatible with
0. On the contrary, the NALPL models clearly need some
positive luminosity evolution to fit the data.

Concerning the $r_d\times H_0/c$ cosmological parameter, we can
observe that the $\Lambda$CDM values are compatible with the no
evolution case, while the values obtained for the NALPL models are
compatible with a lower value. This is also the case for the $H_0$
parameter, as we can see in the last panel.

Focusing on the ability of these models to fit the data, all of them provide a very good fit to the data with a goodness of
fit statistic value of $P(\chi^2,\nu) > 0.9$, as we can see in the
left plot of the bottom panel of
Fig.\,\ref{fig1}. We present the difference of $\chi^2$ values given
by $\Delta \chi^2=\chi^2_{\Lambda\text{CDM}}-\chi^2_{\text{NALPL}}$ in
the right plot of Fig.\,\ref{fig1} bottom panel. It is important to
notice that, in this case, $\Delta \chi^2$ is equal to the difference
of widely used standard model comparison criteria, such as the Akaike
information criterion [\cite{AIC}] or the Bayesian information
criterion [\cite{BIC}], because both $\Lambda$CDM and NALPL have the
same number of free parameters and we are using the same data for the
fits. However, we are only interested in the ability of NALPL to fit
the data, and we are not in search of performing a model comparison against $\Lambda$CDM. In
the plot we also show the standard Jeffrey scale [\cite{Jeffrey}]
to provide a qualitative idea of the strength of the
$\Delta \chi^2$ variation. We consider $0 \leq\Delta\chi^2<1.1$ as a
weak variation (thus compatible $\chi^2$ values), $1.1 \leq
\Delta\chi^2<3$ as a definite variation, $3\leq \Delta\chi^2 <5$ as a
strong variation, and $5 \leq \Delta\chi^2$ as a very strong variation.

From these results, we can observe that most
NALPL models (A1, B1, B2, C1, C2, C3, D2, and D3) are not only able to fit
the data with a very high goodness of fit statistic, but their
$\chi^2$ value is also
compatible with that obtained for $\Lambda$CDM.

In Fig.\,\ref{fig2} we present the results obtained when adding the
measurements of the growth of matter perturbations to the low-redshift
background probes. In the top panel we show the best-fit values for the cosmological and nuisance
parameters. The obtained values for the $\Lambda$CDM models are
completely compatible with the no luminosity evolution case, as in the
previous case (Fig.\,\ref{fig1}). The only difference here are
slightly smaller error bars due to the introduction of more data
points. Concerning the NALPL models, we have an
extra cosmological parameter, $\Omega_m$, which is very well
constrained, but the other parameters remain qualitatively compatible
with the results from Fig.\,\ref{fig1}: the cosmological parameters
$n$, $r_d\times H_0/c$ and $H_0$ are compatible
with lower values than for the no luminosity evolution case, while
$\epsilon$ is clearly not compatible with 0.

In the bottom panel of Fig.\,\ref{fig2} we provide the results for the
goodness of fit statistic and the variation of the $\chi^2$ values. In the first plot
we can observe that all models provide a very good fit to the data
($P(\chi^2,\nu)>0.93$), while in the second panel we show that most
NALPL models (A1, B1, B2, C1, C2, C3, D2, and D3) have a $\chi^2$ value
compatible with (or slightly better than) those provided
by $\Lambda$CDM. From a model criteria point of view it is clear that
NALPL is slightly disfavored because of the
introduction of the extra $\Omega_m$ parameter. However, the
importance of the Occam factor depends on the model criteria used and,
as discussed before, our goal is not to test the NALPL model against
$\Lambda$CDM to adopt this model as an alternative, but just to show that it can fit the observational data
extremely well.

In Fig.\,\ref{fig3}, just for
completeness and illustrative purposes, we present the prediction for
all probes using
the best low-redshift power law model (B2), $\Lambda$CDM
B2, and $\Lambda$CDM imposing no luminosity evolution. We used the global
best-fit values for the cosmological and nuisance
parameters. It is clear that all three models are able to reproduce the observations extremely well.

Taking all the low-redshift probes into account (SNIa, BAO, $H(z),$ and
$f\sigma_8(z)$), a NALPL model is
perfectly compatible with the data for several luminosity evolution models. This points to the fact that
low-redshift probes do not definitively prove the acceleration of the
Universe and that we need more precise low-redshift measurements to
claim this acceleration firmly.

%-----------------------------------------------------------------

\section{Conclusions}\label{sec6}
In this work we have analyzed the ability of the low-redshift probes
(SNIa, BAO, $H(z)$, and $f\sigma_8(z)$) to prove the accelerated expansion of the
Universe. More precisely, we considered a
nonaccelerating low-redshift power law cosmology and checked its
ability to fit these cosmological data. Using only the low-redshift background probes, and not imposing the SNIa
intrinsic luminosity to be redshift independent (accounting for
several luminosity evolution models), we find that a nonaccelerated
low-redshift power law cosmology is able to fit very well all the
observations, for all the luminosity evolution models
considered. Moreover, most of the NALPL models provide a value for the $\chi^2$ perfectly
compatible with that obtained for $\Lambda$CDM.

When we add the measurements of the growth of matter perturbations, a
nonaccelerated low-redshift power law cosmology is able to fit
all
the data extremely well  for all the luminosity evolution models considered. As in
the previous case, most of the NALPL models provide $\chi^2$ values
that are perfectly compatible (or even slightly better) than those
coming from $\Lambda$CDM.

The main conclusion of this work is that if we do not impose the SNIa
intrinsic luminosity to be independent of the redshift, the
combination of low-redshift probes is not sufficient to firmly prove the
accelerated expansion of the Universe.

%\begin{acknowledgements}

%\end{acknowledgements}


%\bibliographystyle{aa}
%\bibliography{aa_v2}
%\end{document}

\end{document}